\documentclass[iop]{emulateapj}              
\newcommand{\blandscape}{\begin{turnpage}} 
\newcommand{\elandscape}{\end{turnpage}} 
  
\usepackage{graphicx,fancyvrb,epstopdf,xcolor,booktabs,ifthen,subfigure,captcont} 
\usepackage{verbatim,catchfilebetweentags,longtable,rotating,pstricks-add}
\usepackage{changepage,multirow,upgreek,natbib,gnuplottex,bm,amsmath}
\usepackage[utf8x]{inputenc}
\usepackage[flushleft]{threeparttable}
\usepackage[colorlinks=true,citecolor=blue]{hyperref}
\usepackage[final]{pdfpages}
\usepackage[nomessages]{fp}
\usepackage[mathscr]{euscript}
\usepackage{subfiles}
\usepackage{enumitem}
\usepackage{chngcntr}
\def\deg{\hbox{$^\circ$}}              
\def\h{\hbox{$^{\rm h}$}}              
\def\ergs{$\rm{erg}~\rm{s}^{-1}$}       
\def\lx{$\rm{L}_{X}$}                   
\def\lbol{$\rm{L}_{bol}$}                   
\def\lxs{$\rm{L}_{X,[0.3-10]}$}                 
\def\lxh{$\rm{L}_{X,[14-50]}$}                 
\def\sun{$_{\odot}$}                  
\def\zsun{$\rm{Z}_{\odot}$}           
\def\rstar{$\rm{R}_{\star}$}          
\def\cts{counts s$^{-1}$}
\def\td{$\tau_{d}$}
\def\tr{$\tau_{r}$}

\def\nh{$\rm{N_{H}}$}
\def\h1{{\sc Hi}}
\def\xmm{{\it XMM-Newton}}
\def\swift{{\it Swift}}
\def\chandra{{\it Chandra}}
\def\rosat{{\it ROSAT}}

\def\apec{{\sc apec}}
\def\xspec{{\sc xspec}}
\def\chisq{$\chi^2$}
\newcommand{\Ka}{K$\alpha$}
\newcommand{\E}[1]{$\times 10^{#1}$}
\newcommand{\Pten}[1]{$10^{#1}$}

\newcommand{\scz}{\scriptsize}

\newcommand{\addcol}[1]{
\hline
\end{tabular}
& 
\begin{tabular}{#1}
\hline
}

\newcommand{\tabstart}{
\begin{table}
\begin{threeparttable}
}

\newcommand{\tabularstart}[2]{
\begin{tabular}{#1} 
  \begin{tabular}{@{}#2}
\hline
}
\newcommand{\tabularend}{
\hline
\end{tabular}
\end{tabular} 
}

\newcommand{\tabnote}{
\begin{tablenotes}
}

\newcommand{\tabend}{
\end{tablenotes}
\end{threeparttable}
\end{table}
}

\newcommand{\RN}[1]{%
  \textup{\uppercase\expandafter{\romannumeral#1}}%
}

\DeclareRobustCommand*{\ion}[2]{%
  #1$\;$%
  \if b\expandafter\@car\f@series\relax\@nil
    \begingroup 
      \sbox0{\rmfamily\mdseries\textsc{v}}%
      \resizebox{!}{\ht0}{\rmfamily\RN{#2}}%
    \endgroup
  \else
    \textsc{\rmfamily\RN{#2}}%
  \fi
}
\bibpunct{(}{)}{;}{a}{,}{,}

\slugcomment{Accepted in The Astrophysical Journal, 2017}

\shorttitle{CC Eridani}
\shortauthors{Karmakar et al.}

\begin{document}

\title{X-ray Superflares on CC Eri}

\author{Subhajeet Karmakar$^{1,2}$\altaffilmark{*}, J. C. Pandey$^{1}$, V. S. Airapetian$^{3}$, and Kuntal Misra$^{1}$}
\affil{$^{1}$ Aryabhatta Research Institute of observational sciencES (ARIES), Nainital 263002, India}
\affil{$^{2}$ Pt.  Ravishankar Shukla University, Raipur 492010, India}
\affil{$^{3}$ NASA Goddard Space Flight Center, 8800 Greenbelt Road, Greenbelt, Maryland 20771, USA}

\altaffiltext{*}{subhajeet@aries.res.in}
\begin{abstract}  
	We present an in-depth study of two superflares (F1 and F2) detected on an active binary star CC Eridani by the \swift\ observatory.  These superflares triggered the Burst Alert Telescope (BAT) in the hard X-ray band on 2008 October 16 and 2012 February 24. The rise phases of both the flares were observed only with BAT, whereas the decay phases were observed simultaneously with the X-ray Telescope. It has been found that the flares decay faster in the hard X-ray band than in the soft X-ray band. Both  flares F1 and F2 are  highly energetic with respective peak X-ray luminosities of $\sim$$10^{32.2}$ and $\sim$$10^{31.8}$ erg s$^{-1}$ in 0.3--50 keV energy band,  which are larger than any other flares previously observed on CC Eri. 
	The time-resolved spectral analysis during the flares shows the variation in the coronal temperature, emission measure, and abundances.  The elemental abundances are enhanced by a factor of $\sim$8 to the minimum observed in the post-flare phase for the flare F1. The observed peak temperatures in these two flares are found to be 174 MK and 128 MK.  Using the hydrodynamic loop modeling, we derive loop lengths for both the flares as 1.2$\pm$0.1$\times$10$^{10}$ cm and 2.2$\pm$0.6$\times$10$^{10}$ cm, respectively.
  The Fe \Ka\ emission at 6.4 keV is also detected in the X-ray spectra and we model the \Ka\ emission feature as fluorescence from the hot flare source irradiating the photospheric iron.  These superflares are the brightest, hottest, and shortest in duration  observed thus far on CC Eri. 
\end{abstract}


\section{Introduction}
\label{sec:intro}
Flares on the Sun and solar-type stars are generally interpreted as a rapid and transient release of magnetic energy in coronal layers driven by reconnection processes, associated with electromagnetic radiation from radio waves to $\gamma$-rays. As a consequence, the charge particles are accelerated and gyrate downward along the magnetic field lines, producing synchrotron radio emission, whereas these electron and proton beams collide with the denser material of the chromosphere and emit in hard X-rays ($>$20 keV). Simultaneous heating of plasma up to tens of MK evaporates the material from the chromospheric footpoints, which in turn increases the density on newly formed coronal loops emitting at extreme UV and X-rays. 
Since the non-flaring coronal emission only contains the information about an optically thin, multi-temperature, and possibly multi-density plasma in coronal equilibrium; therefore, it is very important to understand the dynamic behavior of the corona flaring events.   
Extreme flaring events are even more useful to understand the extent to which the dynamic behavior can vary within the stellar environments.

The typical total energy of solar flares ranges from \Pten{29-32} erg, whereas the flares on normal solar-type stars, having a total energy range of \Pten{33-38} erg, are generally termed as ``superflares'' \citep{Schaefer-00-23,Shibayama-13-2}. In the past, X-ray superflares have been observed and analyzed in the late-type stars Algol \citep{Favata-99-1}, AB Dor \citep{Maggio-00}, EV Lac \citep{Favata-00-6, Osten-10-5}, UX Ari \citep{Franciosini-01-1}, II Peg \citep{Osten-07-3}, and DG CVn \citep{Fender-15-5}.
Recent observations of X-ray  superflares reveal an important aspect of the detection of iron \Ka\ fluorescence emission during flaring events \citep{Ercolano-08-1,Testa-08-6}. Previously, this iron \Ka\ fluorescence emission line was detected on the solar flares \citep{Parmar-84-1,Zarro-92-1}.
The X-ray emissions during the solar and stellar flares interact with the photospheric layers and become reprocessed through scattering and photoionization. These processes also produce characteristic fluorescent emission from astrophysically abundant species.

In this paper, we describe  two superflares serendipitously observed by the \swift\ observatory on an active spectroscopic binary star CC Eri. This binary consists of a K7.5Ve primary and M3.5Ve secondary \citep{Amado-00-1}, and is located at a distance of $\sim$11.5 pc. With a mass ratio $\approx$2 \citep{Evans-59-11}, the system is tidally locked and the primary component is one of the fastest rotating K dwarfs in the solar vicinity with a rotation period of 1.56 day. 
\cite{Demircan-06} estimated the age of the CC Eri system to be 9.16 Gyr.
The chromospheric emission was found to vary in anti-phase with its optical continuum, suggesting the association  of active emission regions with starspots \citep{Busko-77-1,Amado-00-1}. 
The polarization of the quiescent radio emission was found to be 10--20\% \citep{Osten-02-4, Slee-04}, which implies the presence of a large-scale magnetic field.
 The first X-ray detection was done with \textit{HEAO1} showing X-ray luminosity (\lx) of \Pten{29.6} \ergs\ in the 2--20 keV energy band \citep{Tsikoudi-82}. Later several X-ray observations were made with other satellites such as \textit{Einstein} IPC \citep{Caillault-88-3} and \textit{EXOSAT} \citep{Pallavicini-88-1}. Using \xmm\ observations, the quiescent state coronae of CC Eri were well described by two-temperature plasma models ($\sim$3 and $\sim$10 MK) with a luminosity of $\sim$\Pten{29} \ergs\ in 0.3--10 keV energy band \citep{Pandey-08-4}. CC Eri has a record of frequent flaring activity observed across a wide range of the electromagnetic spectrum. The first ever X-ray flare on CC Eri was detected with the \rosat\ satellite \citep{Pan-95-9}. Later several X-ray flares were detected with \xmm\ \citep{Crespo-Chacon-07, Pandey-08-4},  \chandra\ \citep{Nordon-07}, \swift\ \citep{Evans-08-17, Barthelmy-12-56}, and  \textit{MAXI} GCS \citep{Suwa-11-1}.  Among the previously observed flares, the largest one was observed with \chandra, having a peak X-ray flux that is $\sim$11 times more than that of the quiescent value. 
Superflares from CC Eri are expected to be driven by strong surface magnetic fields as estimated by \cite{Bopp-73-4}.

The paper is organized as follows. Observations and the data reduction procedure are discussed in \S~\ref{sec:obs}. Analysis and results from X-ray timing and spectral analysis along with the time-resolved spectroscopy, loop modeling, and Fe \Ka\ modeling are discussed in \S~\ref{sec:result}. 
Finally, in \S~\ref{sec:discussions}, we discuss our results and present conclusions.

\section{Observations and data reduction}
\label{sec:obs}

\subsection{BAT Data }
\label{subsec:obs_bat}
The flare F1 triggered \swift's Burst Alert Telescope \citep[BAT;][]{Barthelmy-05-4} on  2008 October 16 UT 11:22:52 (=$T0_{1}$) during a preplanned spacecraft slew. The flare F2 was detected as an Automatic Target triggered on board on 2012 February 24 UT 19:05:44 (=$T0_{2}$).  
We have used BAT pipeline software within FTOOLS\footnote{The mission-specific data analysis procedures are  provided in FTOOLS software package; a full description of the procedures mentioned here can be found at https://heasarc.gsfc.nasa.gov/docs/software/ftools/ftools\_menu.html} version 6.20 with the latest CALDB version `BAT (20090130)' to 
correct the energy from the efficient but slightly non-linear on board energy assignment.
BAT light curves were extracted using the task {\sc batbinevt}. For the spectral data reported here, the mask-weighted spectra in the 14--50 keV band were produced using {\sc batmaskwtevt} and  {\sc batbinevt} tasks with an energy bin of 80 channels. The BAT ray tracing columns in spectral files were updated using the {\sc batupdatephakw} task, whereas the systematic error vector was applied to the spectra from the calibration database using the {\sc batphasyserr} task. The BAT detector response matrix was computed using the {\sc batdrmgen} task. 
The sky images in two broad energy bins were created  using {\sc batbinevt} and {\sc batfftimage}, and flux at the source position was found using {\sc batcelldetect}, after removing a fit to the diffuse background and the contribution of bright sources in the field of view. The spectral analysis of all the BAT spectra was done using the X-ray spectral fitting package \citep[\xspec; version 12.9.0n;][]{Arnaud-96-6}.
All the errors associated with the fitting of the BAT spectra were calculated for a confidence interval of 68\% ($\Delta\chi^2$ = 1).

\begin{figure*}[ht!]
    \center
  \vspace{-0.85cm}
  \subfigure[Observations in 2008 shows flare F1 with a post-flare (PF) region]{\includegraphics[height=9.3cm,angle=0,trim={0 -0cm 0 0},clip]{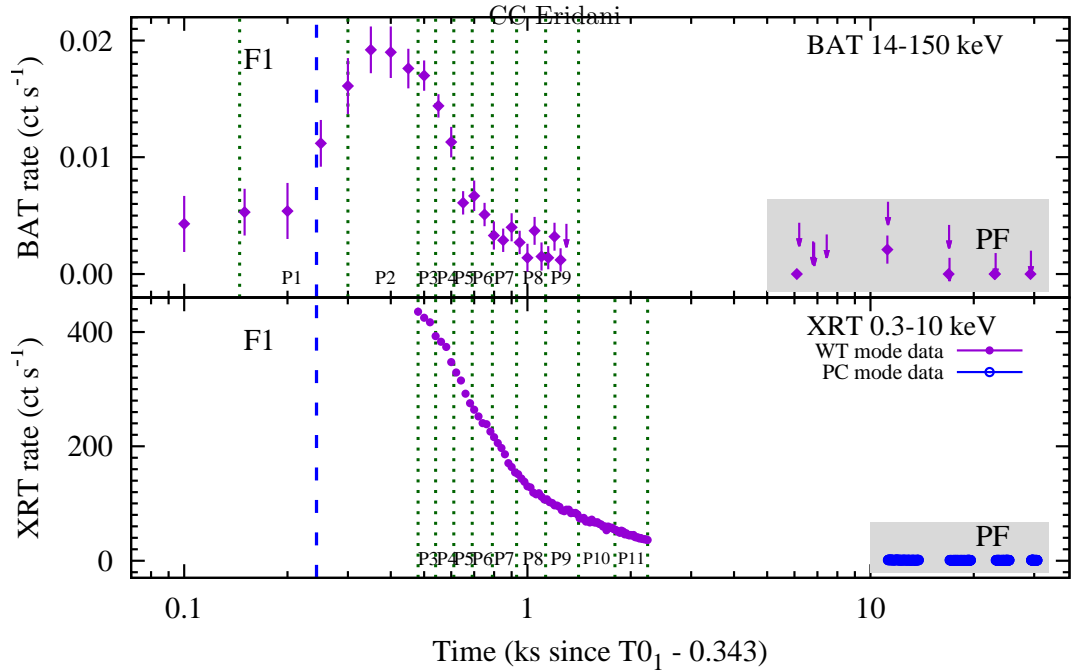}}
  \subfigure[Observations in 2012 shows flare F2]{\includegraphics[height=9.3cm,angle=0]{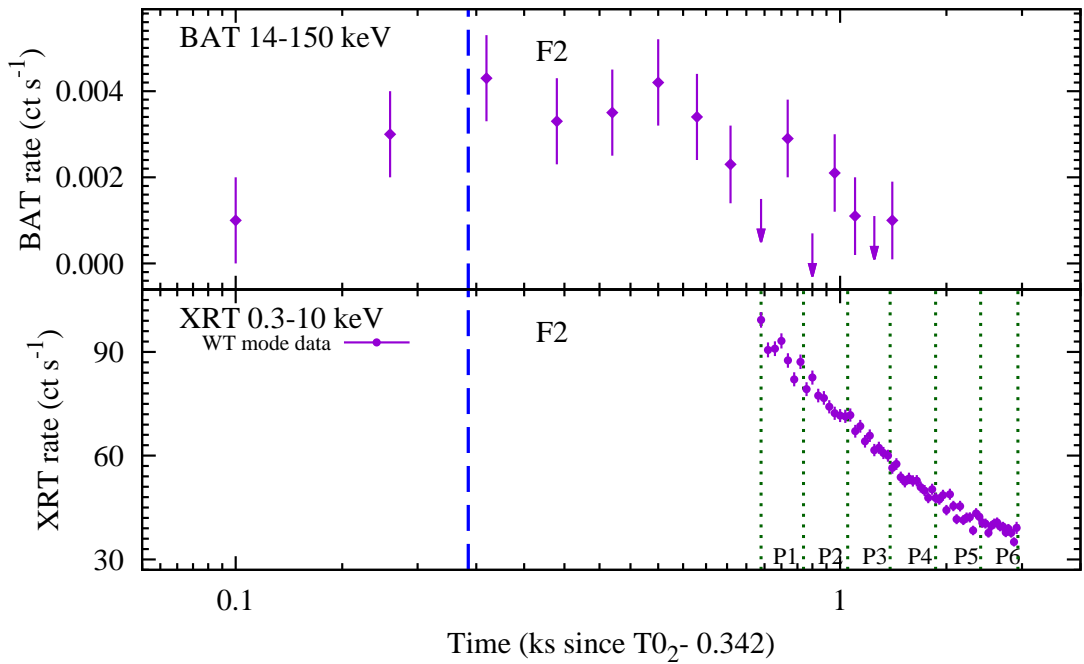}}
	\caption{X-ray light curves of the flares (a) F1 and (b) F2 of CC Eri. Temporal binning for BAT light curves is 50 and 80 s for flares F1 and F2, respectively, whereas the binning for XRT light curves is 10 s for both flares. Dashed vertical lines indicate the trigger time, whereas dotted vertical lines show the time intervals for which time-resolved spectroscopy was performed. The time intervals are represented by $P_i$, where $i$ = 1  -- 11 for F1 and 1-- 6 for F2.}
\label{fig:lc}
\end{figure*}

\begin{figure*}[!htp]
  \center
  \begin{center}
\includegraphics[width=11cm,angle=-90]{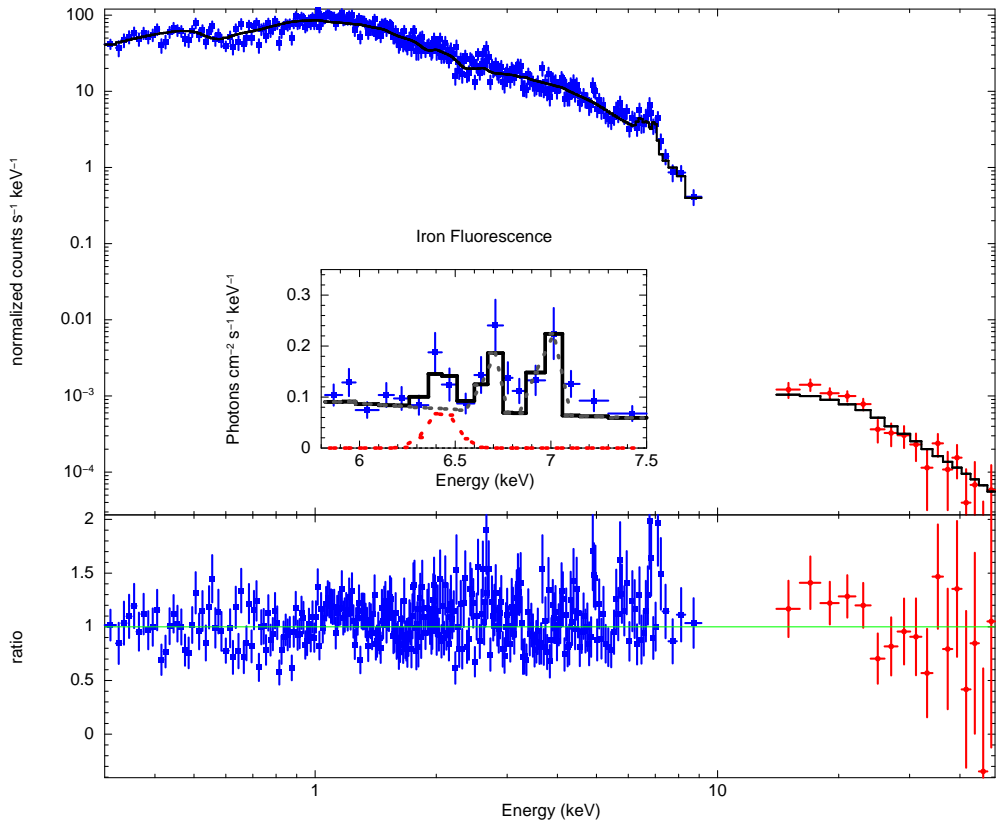}
\caption{Combined XRT and BAT spectra near the peak phase of flare F1 (i.e. part P3) are shown as representative spectra. In the top panel, the XRT and BAT spectra are shown with solid squares and solid circles, respectively, whereas the best-fit \apec\ 3-T model is over-plotted with a continuous line. The bottom panel plots the ratio between data and model. The inset of the top panel shows a close-up view of the Fe \Ka\ complex, where the 6.4 keV emission line is fitted with a Gaussian. The contribution of \apec\ component and the Gaussian component is shown by dashed lines.}
\label{fig:spec}
  \end{center}
\end{figure*}

\renewcommand{\thesubfigure}{(\roman{subfigure})}[t]
\begin{figure*}
  \subfigure[Flare F1]{\includegraphics[width=9.1cm,angle=0]{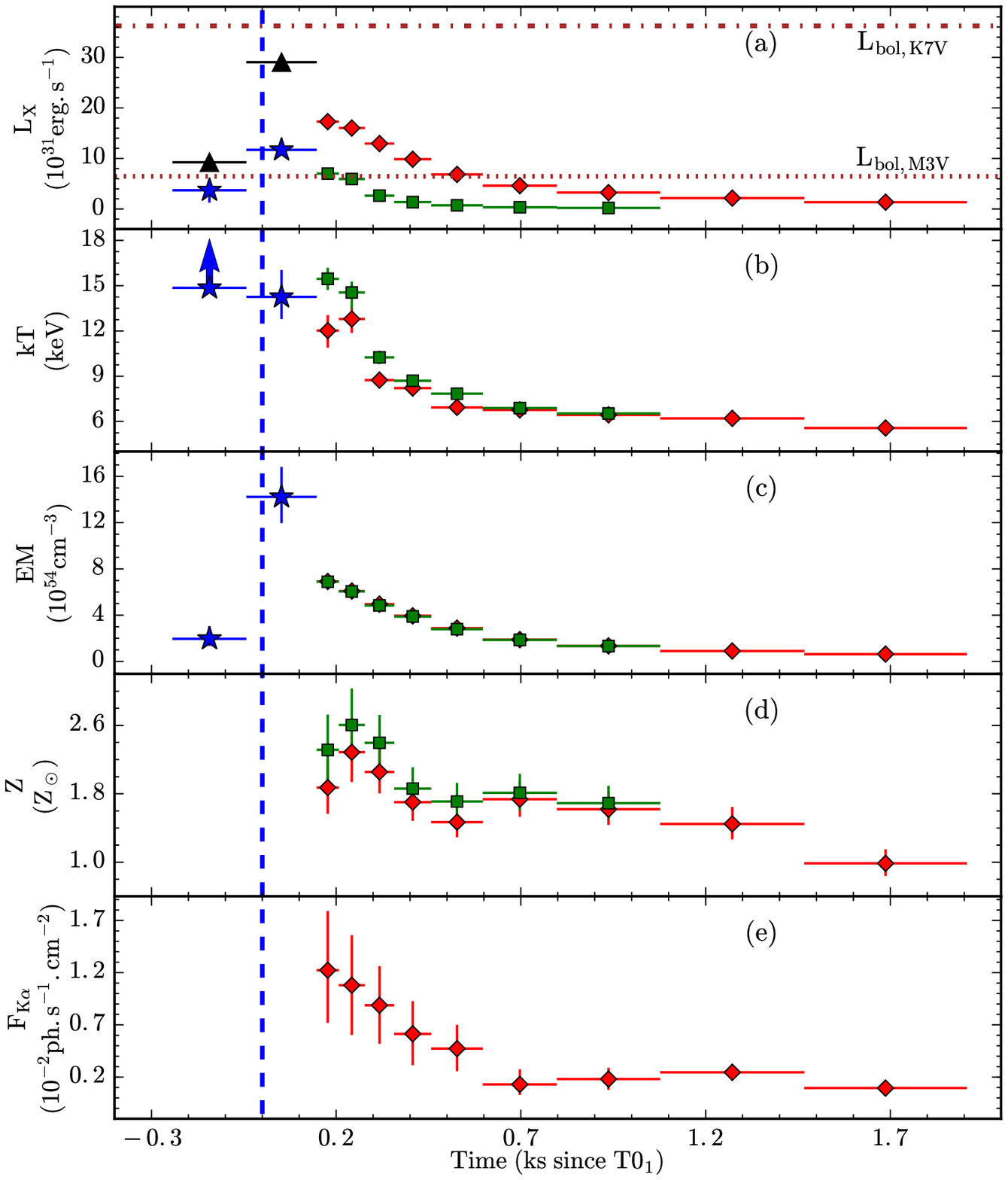}} 
  \subfigure[Flare F2]{\includegraphics[width=9.1cm,angle=0]{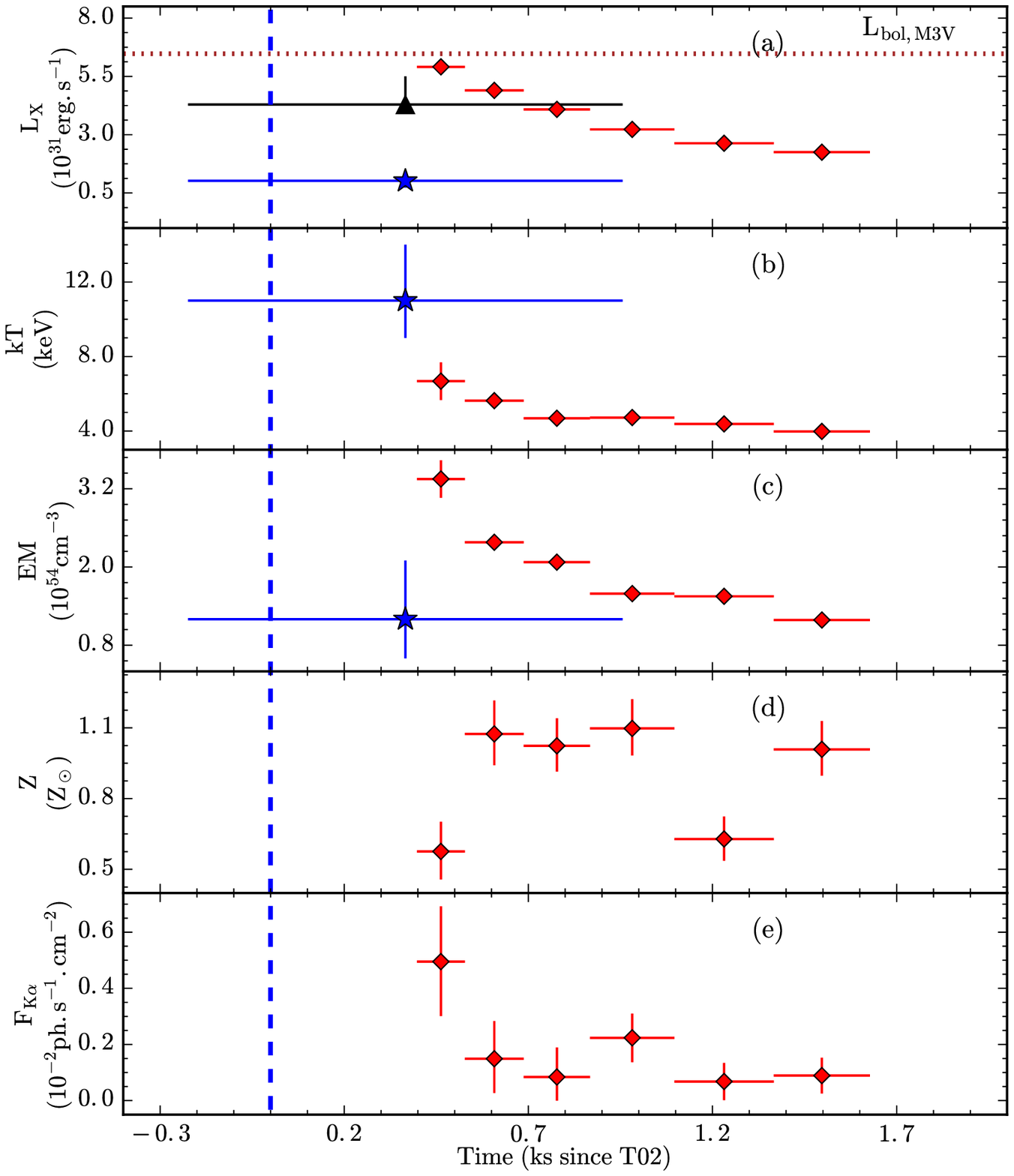}} 
\caption{
  Evolution of spectral parameters of CC Eri during the flares F1 (i) and F2 (ii).
  Parameters derived with the XRT, BAT, and XRT+BAT spectral fitting in all the panels are represented by the solid diamonds, solid stars, and solid squares, respectively. 
  In the top panel (a), the X-ray luminosities are derived in 0.3--10 keV (solid diamond) and 14--50 keV (solid star and solid squares) energy bands. For the first two segments, BAT luminosity derived in the 14--50 keV energy band is extrapolated to the 0.3--10 keV energy band (solid triangles). The dashed--dotted and dotted horizontal lines correspond to bolometric luminosity of the primary and secondary components of CC Eri, respectively.     
  Panels (b)--(e) display the variations of plasma temperature, EM, abundance, and Fe \Ka\ line flux, respectively.
  The dashed vertical line indicates the trigger time of the flares F1 and F2.
 Horizontal bars give the time range over which spectra were extracted; 
 vertical bars show a 68\% confidence interval of the parameters.}
\label{fig:trs_batxrt}
\end{figure*}
\renewcommand{\thesubfigure}{\alph{subfigure}}

\subsection{X-Ray Telescope (XRT) Data }
\label{subsec:obs_xrt}
Flaring events F1 and F2 were observed by the X-ray telescope  \citep[XRT;][]{Burrows-05-3} from  $T0_{1}$+147.2 and $T0_{2}$+397.5 s, respectively. The XRT observes in the energy range of 0.3--10 keV using CCD detectors, with the energy resolution of $\approx$ 140 eV at the Fe K (6 keV) region as measured at launch time. In order to produce the cleaned and calibrated event files, all the data were reduced using the \swift ~{\sc xrtpipeline} task (version 0.13.2) and
calibration files from the latest CALDB version `XRT (20160609)' release\,\footnote{See {\scz http://heasarc.gsfc.nasa.gov/docs/heasarc/caldb/swift/}}. The cleaned event lists generated with this pipeline are free from the effects of hot pixels and the bright Earth. 
In the case of flare F1, due to the large XRT count rate, the initial data recording was in Windowed Timing (WT) mode; whereas from  $T0_{1} +11.7$ s until the end of the observation ( $T0_{1}+31.1$ s), data were taken in Photon Counting (PC) mode. The flare F2 was observed only in WT mode after $T0_{2}$ for 1.2 ks.
From the cleaned event list, images, light curves, and spectra for each observation were obtained with the {\sc xselect} (version 2.4d) package. We have used only grade 0--2 events in WT mode and grade 0--12 events in PC mode to optimize the effective area and hence the number of collected counts.
Taking into account the point-spread function correction \citep[PSF;][]{Moretti-05-2} as well as the exposure map correction, the ancillary response files for the WT and PC modes were produced using the task {\sc xrtmkarf}. In order to perform the spectral analysis, we have used the latest response matrix files \citep{Godet-09-3}, i.e. {\sc swxwt0to2s6\_20010101v015.rmf} for the flare F1 and {\sc swxwt0to2s6\_20110101v015.rmf} the flare F2 in WT mode and {\sc swxpc0to12s6\_20010101v014.rmf} for flare F1 in PC mode. All XRT spectra were binned to contain more than 20 counts bin$^{-1}$. 
 The spectral analysis of all the XRT spectra was carried out in an energy range of 0.3--10 keV using \xspec. All the errors of XRT spectral fitting were estimated with a 68\% confidence interval ($\Delta\chi^2$ = 1), equivalent to $\pm 1\sigma$.
 In our analysis, the solar photospheric  abundances (\zsun) were adopted from \cite{Anders-89-3}, whereas, to model \nh, we used the cross-sections obtained by \cite{Morrison-83-8}.

 While extracting the light curve and spectra, we took great care in order to correct the data for the effect of pile-up in both WT and PC modes. At high observed count rates in WT mode (several hundred counts s$^{-1}$) and PC mode (above 2 counts s$^{-1}$), the effects of pile-up are observed as an apparent loss of flux, particularly from the center of the PSF and a migration from 0--12 grades to higher grades and energies at high count rates. To account for this effect, the source region of WT data were extracted in a rectangular 40 $\times$ 20 pixel region (40 pixels long along the image strip and 20 pixels wide; 1 pixel = 2$.\!\!^{\prime\prime}$36) with a region of increasing size (0 $\times$ 20--20 $\times$ 20 pixels) excluded from its center, whereas the background region was extracted as 40 $\times$ 20 pixel region in the fainter end of the image strip.
We produced a sample of grade ratio distribution using background-subtracted source event lists created in each region. The grade ratio distribution for the grade 0 event is defined as the ratio of the grade 0 event over the sum of grade 0--2 events per energy bin in the 0.3--10 keV energy range. 
Comparing the grade ratio distribution with that obtained using non-piled-up WT data, in order to estimate the number of pixels to exclude, we find that an exclusion of the innermost 5 pixels for F1 and 3 pixels for F2 were necessary when all the WT data are used. 
In order to carry out a more robust analysis for pile-up corrections, we also fit the spectra with an absorbed power law. The hydrogen column density $N_{\rm H}$ was fixed to the values that were obtained from the fit of the non-piled-up spectrum (exclusion of innermost 20 pixels was assumed to be unaffected by piled-up). This gives similar results of exclusion of the innermost pixels within its 1$\sigma$ value. 
In PC mode, since the pile-up affects the center of the source radial PSF, we fit the wings of the source radial PSF profile with the XRT PSF model \citep[a King function; see][]{Moretti-05-2} excluding 15 pixels from the center and then extrapolated to the inner region. The PSF profile of the innermost 4 pixels was found to deviate from the King function, the exclusion of which enables us to mitigate the effects of pile-up from PC mode data. 


\section{Results and Analysis}
\label{sec:result}

\subsection{X-Ray Light Curves}
\label{sec:xray-lc}
X-ray light curves of CC Eri obtained in 0.3--10 keV and 14--150 keV energy bands are shown in Fig.~\ref{fig:lc}. The BAT observation in 2008, which began at $T0_{1}$--243 s, shows a rise in intensity up to $T0_{1}$+100 s, where it reached a peak intensity with a count rate of 0.024$\pm$0.005 \cts, which is $\sim$24 times higher than the minimum observed count rate. A sharp decrease in intensity up to nearly $T0_{1}$+450 s was observed followed by a shallower decay until the end of the BAT observations at $\sim$$T0_{1}$+950 s. The XRT count rate of flare F1 was already declining as it entered the XRT field of view at $T0_{1}$+142 s. 
The peak XRT count rate for the pile-up corrected data was found to be 437$\pm$7 \cts, which was $\sim$100 \cts ~lower than the previously reported count rate \citep{Evans-08-17}. 
Flare F1 survived for  the entire  WT mode of observation of 1.8 ks, after that the observation was switched over in PC mode where count rates were found to be constant and marked as `PF' in Fig.~\ref{fig:lc}(a).
The BAT observation of flare F2, which began at $T0_{2}$--240 s, shows a rise in intensity and peaked around $T0_{2}$, followed by a decline in intensity until the end of the BAT observation.
The peak BAT count rate for flare F2 was found to be $\sim$0.004 \cts, which was $\sim$4 times more than the minimum count rate observed.  The flare `F2' was also already declining when it entered in the XRT field of view with a pile-up corrected peak count rate of 99$\pm$2 \cts. The XRT count rates at the peak of the flares F1 and F2 were found to be 474 and 108 times more than the minimum observed count rates, respectively. The durations of the flares  were  more than 2.2 for F1  and 1.8 ks for F2. 
The e-folding rise times (\tr) of the flares F1 and F2 derived from BAT data were found to be 150$\pm$12 and 146$\pm$34 s, respectively; whereas e-folding decay times (\td) with BAT and XRT data were found to be  283$\pm$13 and 539$\pm$4 s for flare F1 and 592$\pm$114 and 1014$\pm$17 s for flare F2, respectively. This indicates a faster decay in the hard X-ray band than in the soft X-ray.
Both flaring events were clearly identified in the energy band of 14--50 keV, but they were barely detectable above 50 keV.

\subsection{BAT Spectral Analysis}
\label{sec:BAT}
Spectral parameters during the flares evolve with time as flare emission rises, reaches its peak, and later decays similarly to the X-ray light curve. Therefore, in order to trace the spectral change, we have divided the BAT light curve of the flare F1 into nine time segments and extracted the spectra of each segment, whereas we could not divide the flare F2 into any further segments due to poor statistics.
The combined BAT and XRT spectra near the peak phase of the flare F1 (i.e. part P3)  of CC Eri are shown in Fig.~\ref{fig:spec}.
The dotted vertical lines in the top panel of Fig.~\ref{fig:lc}(a) indicate the time intervals during which BAT spectra were accumulated. 
In this section, we analyze those time segments where only the BAT observations were available, i.e. the first two time segments for flare F1 (P1 and P2), and only one for flare F2.
The hard X-ray spectra were best-fitted using single temperature Astrophysical Plasma Emission Code \citep[\apec; see][]{Smith-01-62} as implemented for collisionally ionized plasma. The addition of another thermal or non-thermal component does not improve the fit-statistics.
Since the standard \apec\ model included in the \xspec\ distribution only considers emission up to photon energies of 50 keV; therefore, we restricted our analysis in 14--50 keV energy band.  The abundances in this analysis were fixed to the mean abundances derived from XRT spectral fitting (see \S~\ref{sec:xray-spectra}) given in a multiple of the solar values of \citet{Anders-89-3}. The galactic \h1 ~column density (\nh) in the direction of CC Eri is calculated according to the survey of \cite{Dickey-90-10} and kept fixed at the value of 2.5\E{20} cm$^{-2}$. The unabsorbed X-ray fluxes were calculated using the {\sc cflux} model.  The variation in hard X-ray luminosity (\lxh), plasma temperature ($kT$), emission measure (EM), and abundances ($Z$) derived from the BAT spectra are illustrated in Fig.~\ref{fig:trs_batxrt} and given in Table~\ref{tab:par_b-xb}. 
The peak plasma temperatures were derived to be $>$14.4 keV and $\sim$11 keV for flares F1 and F2, whereas peak $L_{x,[14-50]}$ were derived to be $1.2\pm0.1 \times 10^{32}$ and $1.1\pm0.1\times10^{31}$ \ergs.

\begin{table}
\tabcolsep=0.4cm
\begin{threeparttable}
\caption{Spectral Parameters Derived for the Post-flare Emission of CC Eri from \swift\ XRT PC mode Spectra During the Time Interval ``PF'' As Shown in Fig.~\ref{fig:lc} Using \apec ~2-T and \apec ~3-T Models.}
\label{tab:par_qs}
\begin{tabular}{lcc}
 \hline\hline
  Parameters    &		2T			&	3T			     \\
 \hline  
\nh\   ($10^{20}$ atoms cm$^{-2}$)	&	$0.9_{-0.8}^{+0.8}$	     	&  	$2.8_{-1.1}^{+1.1}$ \\
 kT$_1$	(keV)				&	$0.83_{-0.02}^{+0.02}$ 		&  	$0.25_{-0.02}^{+0.02}$ \\
 EM$_1$	($10^{52}$ cm$^{-3}$)		&	$3.4_{-0.6}^{+0.6}$ 		&  	$1.4_{-0.4}^{+0.4}$ \\
 kT$_2$	(keV)				&	$3.9_{-0.9}^{+2.4}$ 		&  	$0.94_{-0.03}^{+0.03}$ \\
 EM$_2$	($10^{52}$ cm$^{-3}$)		&	$0.8_{-0.2}^{+0.2}$ 		&  	$1.7_{-0.4}^{+0.4}$ \\
 kT$_3$	(keV)				&	--                           	& 	$3.4_{-0.6}^{+0.7}$ \\
 EM$_3$	($10^{52}$ cm$^{-3}$)		&	--                           	& 	$0.9_{-0.1}^{+0.2}$ \\
 Z	(\zsun)				&	$0.13_{-0.02}^{+0.03}$ 		&  	$0.29_{-0.05}^{+0.09}$ \\
 \lxs\ ($10^{29}$ \ergs)		&	$3.88_{-0.06}^{+0.06}$		&  	$4.33_{-0.07}^{+0.07}$ \\
 \chisq	(dof)				&	  1.56 (120)    		&  	  1.08 (118)           	     \\
 \hline\hline     
\end{tabular}

\tabnote
\item \textbf{Notes.} 
  \nh, kT, and EM are the galactic H{\sc i} column density, plasma temperature, and emission measures, respectively. Z is global metallic abundances and \lxs\ is the derived luminosity in the XRT band.
 \tabend

 \begin{table*}[ht]
\tabcolsep=0.13cm
\begin{threeparttable}
\scriptsize
	\caption{X-Ray Spectral Parameters  of CC Eri During the Flare F1 Derived from the XRT Time-resolved Spectra.}
\label{tab:par_xrt-ka-f1}
\begin{tabular}{cccccccccccccccccccccccc}
\toprule

\multirow{3}{*}{$\mathbf{Parts}$}&
\multirow{2}{*}{$\mathbf{Time~Interval}$}&
\multirow{2}{*}{$\mathbf{kT_3}$}&
\multirow{2}{*}{$\mathbf{EM_3}$}&
\multirow{2}{*}{$\mathbf{Z}$}&
\multicolumn{3}{c}{Fe \Ka}&
\multirow{2}{*}{$\mathbf{L_{x,[0.3-10]}}$}&
\multirow{2}{*}{$\mathbf{\chi^2_{\nu}~(DOF)}$}&
\multirow{2}{*}{$\mathbf{P}$}\\

\cline{6-8}
&&&&&
$\mathbf{E}$&
$\mathbf{EW}$&
$\mathbf{F_{K \alpha}~(10^{-2}}$&&&&&\\
$\mathbf{}$&
$\mathbf{(s)}$&
$\mathbf{(keV)}$&
$\mathbf{(10^{54}~cm^{-3})}$&
$\mathbf{(\rm{Z}_{\odot})}$&
$\mathbf{(keV)}$&
$\mathbf{(keV)}$&
$\mathbf{ph~cm^{-2}~s^{-1})}$&
$\mathbf{(10^{31}~erg~s^{-1})}$&
$\mathbf{}$&
$\mathbf{(\%)}$\\

 \midrule
P3	&  T0$_1$+137:T0$_1$+197    &		$11.6_{-1.1}^{+1.0}$	&		$6.9_{-0.3}^{+0.3}$	&		$1.9_{-0.3}^{+0.3}$	&		$6.37_{-0.08}^{+0.06}$	&		$132_{-67}^{+64}$	&		$1.2_{-0.5}^{+0.6}$	&		$17.3_{-0.2}^{+0.2}$	&		1.179 (303)			&		89.3	\\
P4	&  T0$_1$+197:T0$_1$+267    &		$12.3_{-0.9}^{+0.9}$	&		$6.1_{-0.3}^{+0.3}$	&		$2.3_{-0.4}^{+0.4}$	&		$6.38_{-0.07}^{+0.07}$	&		$124_{-66}^{+46}$	&		$1.1_{-0.5}^{+0.5}$	&		$16.0_{-0.2}^{+0.2}$	&		1.119 (312)			&		88.5	\\
P5	&  T0$_1$+267:T0$_1$+347    &		$8.3_{-0.3}^{+0.5}$	&		$5.0_{-0.2}^{+0.2}$	&		$2.1_{-0.3}^{+0.3}$	&		$6.21_{-0.05}^{+0.05}$	&		$173_{-73}^{+90}$	&		$0.9_{-0.4}^{+0.4}$	&		$13.0_{-0.1}^{+0.1}$	&		0.951 (303)			&		93.9	\\
P6	&  T0$_1$+347:T0$_1$+447    &		$7.7_{-0.4}^{+0.4}$	&		$3.9_{-0.2}^{+0.2}$	&		$1.7_{-0.2}^{+0.2}$	&		$6.34_{-0.05}^{+0.06}$	&		$132_{-77}^{+80}$	&		$0.6_{-0.3}^{+0.3}$	&		$9.9_{-0.1}^{+0.1}$	&		1.149 (292)			&		78.9	\\
P7	&  T0$_1$+447:T0$_1$+587    &		$6.5_{-0.2}^{+0.2}$	&		$2.9_{-0.1}^{+0.1}$	&		$1.5_{-0.2}^{+0.2}$	&		$6.41_{-0.07}^{+0.06}$	&		$77_{-37}^{+92}$	&		$0.4_{-0.2}^{+0.2}$	&		$6.85_{-0.07}^{+0.07}$	&		1.238 (288)			&		80.3	\\
P8	&  T0$_1$+587:T0$_1$+787    &		$6.3_{-0.2}^{+0.2}$	&		$1.88_{-0.08}^{+0.08}$	&		$1.7_{-0.2}^{+0.2}$	&		6.39			&		$23_{-23}^{+25}$	&		$<$0.24			&		$4.62_{-0.05}^{+0.05}$	&		1.134 (282)			&		48.6	\\
P9	&  T0$_1$+787:T0$_1$+1067   &		$6.0_{-0.2}^{+0.2}$	&		$1.34_{-0.05}^{+0.05}$	&		$1.6_{-0.2}^{+0.2}$	&		$6.38_{-0.10}^{+0.08}$	&		$77_{-48}^{+68}$	&		$0.2_{-0.1}^{+0.1}$	&		$3.26_{-0.03}^{+0.03}$	&		1.273 (276)			&		55.1	\\
P10	&  T0$_1$+1067:T0$_1$+1457   &		$5.8_{-0.3}^{+0.3}$	&		$0.90_{-0.04}^{+0.04}$	&		$1.5_{-0.2}^{+0.2}$	&		$6.40_{-0.05}^{+0.05}$	&		$118_{-24}^{+131}$	&		$0.22_{-0.08}^{+0.08}$	&		$2.18_{-0.02}^{+0.02}$	&		1.179 (266)			&		96.4	\\
P11	&  T0$_1$+1457:T0$_1$+1897   &		$5.1_{-0.3}^{+0.2}$	&		$0.63_{-0.03}^{+0.03}$	&		$1.0_{-0.2}^{+0.2}$	&		6.41			&		$68_{-52}^{+55}$	&		$0.06_{-0.05}^{+0.05}$	&		$1.37_{-0.02}^{+0.02}$	&		1.043 (220)			&		77.4	\\
             
\bottomrule
\end{tabular}

\tabnote
\item \textbf{Notes.} 
  kT$_{3}$, EM$_{3}$, and Z are the ``effective'' plasma temperature, emission measures, and abundances during different time intervals of flare decay, respectively. E is the Gaussian peak around 6.4 keV, EW is the equivalent width, F$_{K\alpha}$ is the \Ka\ line flux, \lxs\ is the derived luminosity in XRT band, and P is the F-test probability, which indicates how much of an addition of the Gaussian line at 6.4 keV is significant, i.e. the emission line is not a result of random fluctuation of the data points. All the errors shown in this table are in the 68\% confidence interval.
\end{tablenotes}
\end{threeparttable}
\end{table*}
\normalsize

\blandscape
\center
\begin{table*}[t]
 \begin{threeparttable}
\scriptsize
\tabcolsep=0.08cm
\caption{ Time-resolved Spectral Parameters of CC Eri During the Flare F2 Derived from the XRT Spectra.}
\label{tab:par_xrt-ka-f2}
\begin{tabular}{cccccccccccccccccccccccc}
\toprule

\multirow{3}{*}{$\mathbf{Parts}$}&
\multirow{2}{*}{$\mathbf{Time~Interval}$}&
\multirow{2}{*}{$\mathbf{kT_1}$}&
\multirow{2}{*}{$\mathbf{kT_2}$}&
\multirow{2}{*}{$\mathbf{EM_1}$}&
\multirow{2}{*}{$\mathbf{EM_2}$}&
\multirow{2}{*}{$\mathbf{Z}$}&
\multicolumn{3}{c}{Fe \Ka}&
\multirow{2}{*}{$\mathbf{L_{x,[0.3-10]}}$}&
\multirow{2}{*}{$\mathbf{\chi^2_{\nu}~(DOF)}$}&
\multirow{2}{*}{$\mathbf{P}$}\\

\cline{8-10}
&&&&&&&
$\mathbf{E}$&
$\mathbf{EW}$&
$\mathbf{F_{K \alpha}~(10^{-2}}$&&&&&&&\\
$\mathbf{}$&
$\mathbf{(s)}$&
$\mathbf{(keV)}$&
$\mathbf{(keV)}$&
$\mathbf{(10^{54}~cm^{-3})}$&
$\mathbf{(10^{54}~cm^{-3})}$&
$\mathbf{(\rm{Z}_{\odot})}$&
$\mathbf{(keV)}$&
$\mathbf{(keV)}$&
$\mathbf{ph~cm^{-2}~s^{-1})}$&
$\mathbf{(10^{31}~erg~s^{-1})}$&
$\mathbf{}$&
$\mathbf{(\%)}$\\

 \midrule
 P1	&  T0$_2$+397:T0$_2$+527	&		$1.63_{-0.07}^{+0.05}$	&		$8.7_{-1.0}^{+1.8}$	&		$1.0_{-0.2}^{+0.3}$	&		$2.4_{-0.1}^{+0.1}$	&		$0.54_{-0.12}^{+0.13}$	&		$6.46_{-0.05}^{+0.05}$	&		$181_{-75}^{+108}$	&		$0.50_{-0.19}^{+0.20}$	&		$5.91_{-0.06}^{+0.06}$	&	1.298 (288)	&	92.2	\\
 P2	&  T0$_2$+527:T0$_2$+687	&		$1.25_{-0.06}^{+0.04}$	&		$6.0_{-0.4}^{+0.4}$	&		$0.20_{-0.04}^{+0.05}$	&		$2.18_{-0.07}^{+0.07}$	&		$1.03_{-0.13}^{+0.14}$	&		$6.28_{-0.11}^{+0.13}$	&		$101_{-83}^{+42}$	&		$0.15_{-0.12}^{+0.13}$	&		$4.90_{-0.05}^{+0.05}$	&	1.205 (289)	&	46.2	\\
 P3	&  T0$_2$+687:T0$_2$+867     	&		$1.05_{-0.03}^{+0.02}$	&		$4.9_{-0.2}^{+0.2}$	&		$0.13_{-0.02}^{+0.02}$	&		$1.94_{-0.06}^{+0.06}$	&		$0.98_{-0.11}^{+0.12}$	&		6.35			&		$47_{-47}^{+56}$	&		$0.08_{-0.08}^{+0.11}$	&		$4.09_{-0.04}^{+0.04}$	&	1.164 (285)	&	91.6	\\
 P4	&  T0$_2$+867:T0$_2$+1096	&		$1.02_{-0.03}^{+0.03}$	&		$4.9_{-0.2}^{+0.2}$	&		$0.09_{-0.01}^{+0.01}$	&		$1.50_{-0.05}^{+0.05}$	&		$1.06_{-0.11}^{+0.12}$	&		$6.19_{-0.06}^{+0.06}$	&		$244_{-106}^{+76}$	&		$0.22_{-0.09}^{+0.09}$	&		$3.23_{-0.03}^{+0.03}$	&	0.968 (283)	&	96.7	\\
 P5	&  T0$_2$+1096:T0$_2$+1366   	&		$1.25_{-0.07}^{+0.05}$	&		$4.9_{-0.3}^{+0.4}$	&		$0.21_{-0.05}^{+0.07}$	&		$1.34_{-0.05}^{+0.05}$	&		$0.59_{-0.19}^{+0.09}$	&		6.24			&		$94_{-80}^{+83}$	&		$0.07_{-0.07}^{+0.07}$	&		$2.63_{-0.03}^{+0.02}$	&	1.134 (277)	&	66.4	\\
 P6	&  T0$_2$+1366:T0$_2$+1627 	&		$1.04_{-0.03}^{+0.02}$	&		$4.2_{-0.2}^{+0.2}$	&		$0.08_{-0.01}^{+0.01}$	&		$1.11_{-0.04}^{+0.04}$	&		$0.97_{-0.11}^{+0.12}$	&		$6.25_{-0.13}^{+0.13}$	&		$162_{-114}^{+112}$	&		$0.09_{-0.06}^{+0.06}$	&		$2.25_{-0.02}^{+0.02}$	&	0.925 (253)	&	65.2	\\

\bottomrule
\end{tabular}

\tabnote
\item \textbf{Notes.} 
	Parameters have similar meanings as in Tables \ref{tab:par_qs} and \ref{tab:par_xrt-ka-f1}.
	
\end{tablenotes}
\end{threeparttable}
\end{table*}
\elandscape
\normalsize

\begin{table*}[t]
\tabcolsep=0.28cm
 \begin{threeparttable}
\scriptsize

\caption{Time-resolved Spectral Parameters of the BAT and XRT+BAT Spectra of the Flares F1 and F2}
\label{tab:par_b-xb}
\begin{tabular}{ccccccccccccccccccccccccccc}
\toprule
\multirow{2}{*}{$\mathbf{Flare}$}&
\multirow{2}{*}{$\mathbf{Parts}$}&
$\mathbf{Time~Interval}$&
$\mathbf{kT}$&
$\mathbf{EM}$&
$\mathbf{Z}$&
$\mathbf{F_{7.11-50}~(10^{-2}}$&
$\mathbf{L_{x,[14-50]}}$&
\multirow{2}{*}{$\mathbf{\chi^2_{\nu}~(DOF)}$}\\

$\mathbf{}$&
$\mathbf{}$&
$\mathbf{(s)}$&
$\mathbf{(keV)}$&
$\mathbf{(10^{54}~cm^{-3})}$&
$\mathbf{(\rm{Z}_{\odot})}$&
$\mathbf{ph~cm^{-2}~s^{-1})}$&
$\mathbf{(10^{31}~erg~s^{-1})}$&
$\mathbf{}$\\

 \midrule
 \multirow{9}{*}{F1} 
& P1$^*$& T0$_1$--243 : T0$_1$--43      &		$>14.4$			&		$2.0_{-0.6}^{+1.1}$	&		1.64$^\dagger$		&	--			&		$3.7_{-0.6}^{+0.6}$	&		0.893 (15)		 \\
& P2$^*$& T0$_1$--43 : T0$_1$+137      &		$14_{-1}^{+2}$		&		$14.2_{-2.3}^{+2.6}$	&		1.64$^\dagger$		&	--			&		$11.7_{-1.1}^{+1.0}$	&		0.581 (15)    		 \\
[0.4ex]\cline{2-9}\\ [-2.0ex]                                                                                                                                                                                                                                                    
& P3	& T0$_1$+137 : T0$_1$+197       &		$15.0_{-0.7}^{+0.7}$	&		$6.9_{-0.3}^{+0.3}$	&		$2.3_{-0.4}^{+0.4}$	&		$37.3_{-0.4}^{+0.4}$	&		$7.02_{-0.07}^{+0.07}$		&		1.193 (322)			\\
& P4	& T0$_1$+197 : T0$_1$+267  	&		$14.1_{-1.1}^{+0.7}$	&		$6.1_{-0.3}^{+0.3}$	&		$2.6_{-0.4}^{+0.4}$	&		$33.1_{-0.3}^{+0.3}$	&		$5.95_{-0.06}^{+0.06}$		&		1.104 (331)			\\
& P5	& T0$_1$+267 : T0$_1$+347  	&		$9.80_{-0.4}^{+0.4}$	&		$4.8_{-0.2}^{+0.2}$	&		$2.4_{-0.3}^{+0.3}$	&		$20.3_{-0.2}^{+0.2}$	&		$2.67_{-0.03}^{+0.03}$		&		0.978 (322)			\\
& P6	& T0$_1$+347 : T0$_1$+447  	&		$8.25_{-0.3}^{+0.3}$	&		$3.9_{-0.2}^{+0.2}$	&		$1.9_{-0.2}^{+0.3}$	&		$12.7_{-0.1}^{+0.1}$	&		$1.38_{-0.01}^{+0.01}$		&		1.173 (311)			\\
& P7	& T0$_1$+447 : T0$_1$+587  	&		$7.39_{-0.4}^{+0.4}$	&		$2.8_{-0.1}^{+0.1}$	&		$1.7_{-0.2}^{+0.2}$	&		$7.88_{-0.08}^{+0.08}$	&		$0.752_{-0.008}^{+0.008}$	&		1.256 (307)			\\
& P8	& T0$_1$+587 : T0$_1$+787  	&		$6.44_{-0.2}^{+0.2}$	&		$1.86_{-0.07}^{+0.07}$	&		$1.8_{-0.2}^{+0.2}$	&		$4.58_{-0.05}^{+0.05}$	&		$0.364_{-0.004}^{+0.004}$	&		1.134 (300)			\\
& P9	& T0$_1$+787 : T0$_1$+957  	&		$6.08_{-0.2}^{+0.2}$	&		$1.33_{-0.05}^{+0.05}$	&		$1.7_{-0.2}^{+0.2}$	&		$2.94_{-0.03}^{+0.03}$	&		$0.217_{-0.002}^{+0.002}$	&		1.247 (295)			\\
\midrule
F2$^*$ &	--	&  T0$_2$--242 : T0$_2$+938  &		$11_{-2}^{+3}$		&		$1.2_{-0.6}^{+0.9}$	&		1.00$^\dagger$		&	--		&		$1.1_{-0.1}^{+0.1}$	&	0.683 (15)			\\
\bottomrule
\end{tabular}

\tabnote
\item \textbf{Notes.} 
  kT, EM, and Z are the ``effective'' plasma temperature, emission measures, and abundances during different time intervals of flare decay, respectively.
  F$_{7.11-50}$ is the flux derived in the 7.11--50 keV energy band. \lxh\ is the luminosity derived in the 14--50 keV energy band.
All the errors shown in this table are in the 68\% confidence interval.
\item \textbf{* -} In these time segments, only BAT spectra were available and best-fitted with the single temperature \apec\ model, whereas in other segments, XRT+BAT spectra were fitted with three temperature \apec\ with the first two temperatures fixed to the quiescent value. 
\item \textbf{$^\dagger$ -} The abundances were kept fixed at the average abundance derived from XRT spectral fitting.

 \normalsize
 \end{tablenotes}
\end{threeparttable}
\end{table*}

\subsection{XRT Spectral Analysis}
\label{sec:xray-spectra}
Time-resolved spectral analysis was also performed for the XRT data of both flares. 
The WT mode data for flares F1 and F2 were divided into nine and six time bins, respectively, so that each time bin contains sufficient and a similar number of counts. The length of time bins is variable, ranging from 60--440 s for the flare F1 and 130--261 s for the flare F2. The dotted vertical lines in the bottom panel of Fig.~\ref{fig:lc}(a and b) show the time intervals for which the XRT spectra were accumulated. The first seven time segments of XRT data were common with BAT data for the flare F1.

\subsubsection{Post-flare Phase of the Flaring Event F1 }
\label{subsubsec:qs}

The coronal parameters of the PF phase were derived by the fitting single (1-T), two (2-T), and three (3-T) temperatures \apec ~model. The global abundances ($Z$) and interstellar \h1 column density (\nh) were left as free parameters.
 None of the plasma models (1-T, 2-T, or 3-T) with solar abundances (\zsun) were formally acceptable because large values of $\chi^2$ were obtained.
The 2-T plasma model with sub-solar abundances was found to have a significantly better fit than the 1-T model with reduced $\chi^2$ (\chisq) of 1.56 for 120 degrees of freedom (dof). Adding one more plasma component improves the fit significantly with \chisq\ = 1.08 for 118 dof.
The F-test applied to the $\chi^2$ resulting from the fits with \apec ~2-T and 3-T models showed that the 3-T model was more significant with an F-statistics of 27.7 with a null hypothesis probability of 1.4\E{-10}. The addition of one more thermal component did not show any further improvement in the \chisq; therefore, we assume that the post-flare coronae of CC Eri were well represented by three temperature plasma. Table ~\ref{tab:par_qs} summarizes the best-fit values of the 2-T and 3-T plasma models of various parameters along with their \chisq ~value. 
The first two temperatures in the 3-T model were derived as 0.25$\pm$0.02 and 0.94$\pm$0.03 keV.  These two temperatures are consistent to that derived by \cite{Crespo-Chacon-07} and \cite{Pandey-08-4} for quiescent coronae of CC Eri using \xmm ~data. This indicates that the post-flaring region has not yet returned to the quiescent level and has a third thermal component of $3.4_{-0.6}^{+0.7}$ keV.
With the preliminary analysis of the same data, \cite{Evans-08-17} also suspected that the post-flare region was not a quiescent state.
The X-ray luminosity in the 0.3--10.0 keV band during the post-flare region was derived to be 4.33$\pm$0.07\E{29} erg s$^{-1}$, which was $\sim$5 times higher than the previously  determined quiescent state luminosity in the same energy band by \cite{Pandey-08-4} using \xmm.

\subsubsection{The flaring event F1}
\label{subsubsec:trsxrt}
A 3-T plasma model was found to be acceptable in each segment of flaring event F1. The first two temperatures, corresponding EMs, and \nh\ were found to be constant within a 1$\sigma$ level. 
The average values of all the segments of the first two temperatures were 0.3 $\pm$ 0.1 and 1.1 $\pm$ 0.2 keV, respectively. These two temperatures were very similar to that of the first two temperatures of the PF phase. Therefore, for the further spectral fitting of flare-segments of the flare F1, we fixed the first two temperatures to the average values. The free parameters were temperature and corresponding normalization of the third component along with the abundances.
The time evolution of derived spectral parameters of flare F1 is shown in Fig.~\ref{fig:trs_batxrt}(a) and are given in Table \ref{tab:par_xrt-ka-f1}. The abundance, temperature, and corresponding EM were found to vary during the flare.
The peak values of abundances were derived to be 2.3$\pm$0.4 \zsun , which was $\sim$8 times more than the post-flaring region and $\sim$13 times more than that of the quiescent value of CC Eri \citep{Pandey-08-4}.
The derived peak flare temperature of $12.3\pm0.9$ keV was $\sim$3.6 times more than the third thermal component observed in the PF phase.
The EM followed the flare light curve and peaked at a value of 6.9$\pm$0.3\E{54} cm$^{-3}$, which was $\sim$766 times more than the minimum value observed  at the post-flare region. 
The peak X-ray luminosity in 0.3--10 keV energy band during flare F1 was derived to be \Pten{32.2} \ergs, which was $\sim$400 times more luminous than that of the post-flare regions, whereas $\sim$1922 times more luminous than that of the quiescent state of CC Eri derived by \cite{Pandey-08-4}. The amount of soft X-ray luminosity during the time segments when only BAT data were collected were estimated by extrapolating the 14--50 keV luminosity derived from the  best-fit \apec ~model of the BAT data using {\sc webpimms}\footnote{https://heasarc.gsfc.nasa.gov/cgi-bin/Tools/w3pimms/w3pimms.pl} and  is shown by solid triangles in the top panels of Fig.~\ref{fig:trs_batxrt}(i).

\subsubsection{The flaring event F2}
\label{subsubsec:trsxrt}

For the flaring event F2, no pre-/post-flare or quiescent states were observed; therefore, a time-resolved spectroscopy was done by fitting 1-T, 2-T, and 3-T plasma models. A 2-T plasma model was found suitable for each flare segment as the minimum value of \chisq\ was obtained.
Initially, in the spectral fitting \nh ~was a free parameter and was constant within a 1$\sigma$ level; therefore, in the next stage of spectral fitting, we fixed \nh ~to its average value. The derived spectral parameters are given in Table  \ref{tab:par_xrt-ka-f2}. Both the temperatures and the corresponding EM along with the global abundances  were found to be variable during the flare. In order to compare the plasma properties of the flare  represented by two-temperature components, the total emission measure and temperature were calculated as ${\rm EM = (EM_1+EM_2)}$  and ${\rm T =  (EM_1 .T_1 + EM_2 .T_2)/EM}$. The time evolution of spectral parameters along with the X-ray luminosity in 0.3-10 keV energy band for the flare F2 is shown in Fig.~\ref{fig:trs_batxrt}(ii). The abundances were varied from  0.5 to 1.1 \zsun.
The flare temperature (weighted sum) and total EM  were  peaked at $6.7\pm1.0$ keV and  3.3$\pm$0.3\E{54} cm$^{-3}$, respectively. These values are two to three times higher than the respective minimum observed values. At the end of the flare, the X-ray luminosity was found to be 38\% of its maximum value of \Pten{31.8} \ergs. 

\subsubsection{The emission line at 6.4 keV}
\label{subsec:feline}
In the spectral fitting of XRT spectra with the 3-T \apec ~model for F1 and the 2-T \apec\ model for F2, a significant positive residuals redward of the prominent Fe K complex at $\approx$6.7 keV were detected. This excess emission occurs at the expected position of the 6.4 keV Fe fluorescent line, which is not included in the \apec\ line list. To determine whether indeed such emission is present in the XRT spectrum, we have fitted again the spectra with an additional Gaussian line component along with the best-fit plasma model. Initially, keeping free the width of the Gaussian line ($\sigma$), it converges to a very large value than the actual line width. Therefore, in order to get a best fit, we fixed the $\sigma$ in every value from 10--200 eV with an increment of 10 eV. In this analysis, we have used the $\sigma$ corresponding to the minimum \chisq\ value, ranging from 40--80 eV for a different time segment of the flare. The line centroid (E) and the normalization along with the temperature, abundances, and EM were left free to vary. The best-fit parameters are given in Table \ref{tab:par_xrt-ka-f1} and  \ref{tab:par_xrt-ka-f2}, for the flares F1 and F2, respectively. In each segment, the best-fit line energy agrees with the Fe fluorescent feature at E $\sim$ 6.4 keV (see the sixth column of Table \ref{tab:par_xrt-ka-f1} and eighth column of Table \ref{tab:par_xrt-ka-f2}).
We applied the F-test to the $\chi^2$ resulting from the fits with and without an additional Gaussian line, which shows the significance of the Fe \Ka\ feature with  a probability of the line not being a result of random fluctuation (see the last column of Table \ref{tab:par_xrt-ka-f1} and \ref{tab:par_xrt-ka-f2}).
The derived Fe \Ka\ line flux shows variability and follows the light curve and peaked at a value of $1.2_{-0.5}^{+0.6}$\E{-2} photons s$^{-1}$cm$^{-2}$ for the flare F1 and $5.0_{-1.9}^{+2.0}$\E{-3} photons s$^{-1}$cm$^{-2}$ for the flare F2, which is $\sim$20 and $\sim$7 times more than the minimum observed Fe \Ka\ flux, respectively. The equivalent width (EW) was found to be in the range of 23--173 eV for the flare F1 and 47--244 eV for the flare F2.  

\subsection{XRT+BAT Spectral Analysis}
\label{sec:xrt-bat}
Time-resolved spectroscopy for the  flare F1 was also performed with the XRT+BAT data. Very poor statistics of the BAT spectra did not allow us a time-resolved spectral analysis of the XRT+BAT spectra for the flare F2. A similar approach, as applied for the XRT spectral fitting was also applied for the XRT+BAT spectral fitting.
For the flare F1, we choose seven time bins (P3--P9) similar to the  spectral fitting of only-XRT data as described in the previous section (see Fig.~\ref{fig:lc}). 
Since galactic \h1 column density was not found to be variable during only-XRT spectral analysis; therefore, we fixed \nh ~to its average value. The global abundances, temperature, and corresponding EMs of the third component were free parameters in the spectral fitting.
The derived parameters are given in Table \ref{tab:par_b-xb} and the variations of the spectral parameters are shown in Fig.~\ref{fig:trs_batxrt}. 
The peak temperature derived in this spectral fitting was found to be 15.0$\pm$0.7 keV, which is higher than the highest temperature derived from XRT spectral analysis. The EM was found to have the similar values as those derived from XRT spectral fitting, whereas the peak abundance was found to be $\sim$1.1 times higher than to that derived from XRT data. 

\begin{table}[t]
\tabcolsep=0.12cm
\begin{threeparttable}
\caption{Loop parameters derived for flares F1 and F2}
\label{tab:par_loop}
\tabularstart{l@{}l@{}l@{}l@{}}{l}

\textbf{Sl.} \\
	\hline
1\\
2\\
3\\
4\\
5\\
6\\
7\\
8\\
9\\
10\\
11\\
12\\
13\\
14\\
15\\
16\\
17\\
	\addcol{ll}
\textbf{Parameters} \\
	\hline
$\tau_\mathrm{r,14-150}$ ($\mathrm{s}$)\\
$\tau_\mathrm{d,14-150}$ ($\mathrm{s}$)\\
$\tau_\mathrm{d,0.3-10}$ ($\mathrm{s}$)\\
$L_\mathrm{X, max}$ ($10^{31}~\mathrm{erg~s^{-1}})$ \\
$T_\mathrm{max}$ ($10^{6}~\mathrm{K})$\\
$\zeta$ \\
$L$ ($10^{10}~\mathrm{cm})$ \\
$p$ ($10^{5}~\mathrm{dyn~cm^{-2}})$ \\
$n_\mathrm{e}$ ($10^{12}~\mathrm{cm^{-3}})$ \\
$V$ ($10^{29}~\mathrm{cm^{3}})$ \\
$B$ ($\mathrm{kG})$ \\
$E_\mathrm{H}$ ($10^{3}~\mathrm{erg~s^{-1}~cm^{-3}})$ \\
H ($10^{32}~\mathrm{erg~s^{-1}})$ \\
$E_\mathrm{X, tot}$ ($10^{35}~\mathrm{erg})$ \\
$B_\mathrm{0}$ ($\mathrm{kG})$ \\
$N_\mathrm{loops}$ \\
$\theta(\deg)$ \\
	\addcol{ll}
\textbf{Flare F1} \\
	\hline
150 $\pm$ 12\\ 
283 $\pm$ 13\\ 
539 $\pm$ 4\\ 
17.3 $\pm$ 0.2\\ 
365 $\pm$ 33\\ 
0.550 $\pm$ 0.047\\ 
1.2 $\pm$ 0.1\\ 
15 $\pm$ 5\\ 
15 $\pm$ 7\\ 
$\sim$0.31\\ 
$\sim$6.1\\ 
$\sim$6.6\\ 
$\sim$2.1 \\ 
$>$1.4 \\ 
$>$12 \\ 
$\sim$1\\ 
$\sim$90\\ 
	\addcol{ll}
\textbf{Flare F2} \\
	\hline
146 $\pm$ 34\\  
592 $\pm$ 114\\ 
1014 $\pm$ 17\\ 
5.91 $\pm$ 0.06\\ 
170 $\pm$ 32\\ 
0.819 $\pm$ 0.179 \\ 
2.2 $\pm$ 0.6\\ 
0.8 $\pm$ 0.7\\ 
1.7 $\pm$ 1.7\\ 
$\sim$11.5\\ 
$\sim$1.4\\ 
$\sim$0.13\\ 
$\sim$1.5 \\ 
$>$1.7 \\ 
$>$2 \\ 
$\sim$3\\
--\\
        \tabularend 

\tabnote
\item
1, 2: {e-folding rise and decay times derived from the BAT light curve.}
3: {e-folding decay time derived from the XRT light curve.}
4: {luminosity at the flare peak in the XRT band.}
5: {the maximum temperature in the loop at the flare peak.}
6: {slope in the density-temperature diagram during the flare decay.}
7: {flaring loop length.}
8: {maximum loop pressure at the flare peak.}
9: {maximum electron density in the loop at the flare peak.}
10: {loop volume of the flaring region.}
11: {minimum magnetic field.}
12: {heating rate per unit volume at the flare peak.}
13: {total heating rate at the flare peak.}
14: {total radiated energy.}
15: {total magnetic field required to produce the flare.}
16: {the number of loops needed to fill the flare volume assuming loop aspect ratio of 0.1.}
17: {the astrocentric angle between observer and the flare location on the stellar disk.}
(See the text for a detailed description).
\tabend

\begin{figure*}[t]
\includegraphics[width=6cm,angle=-90]{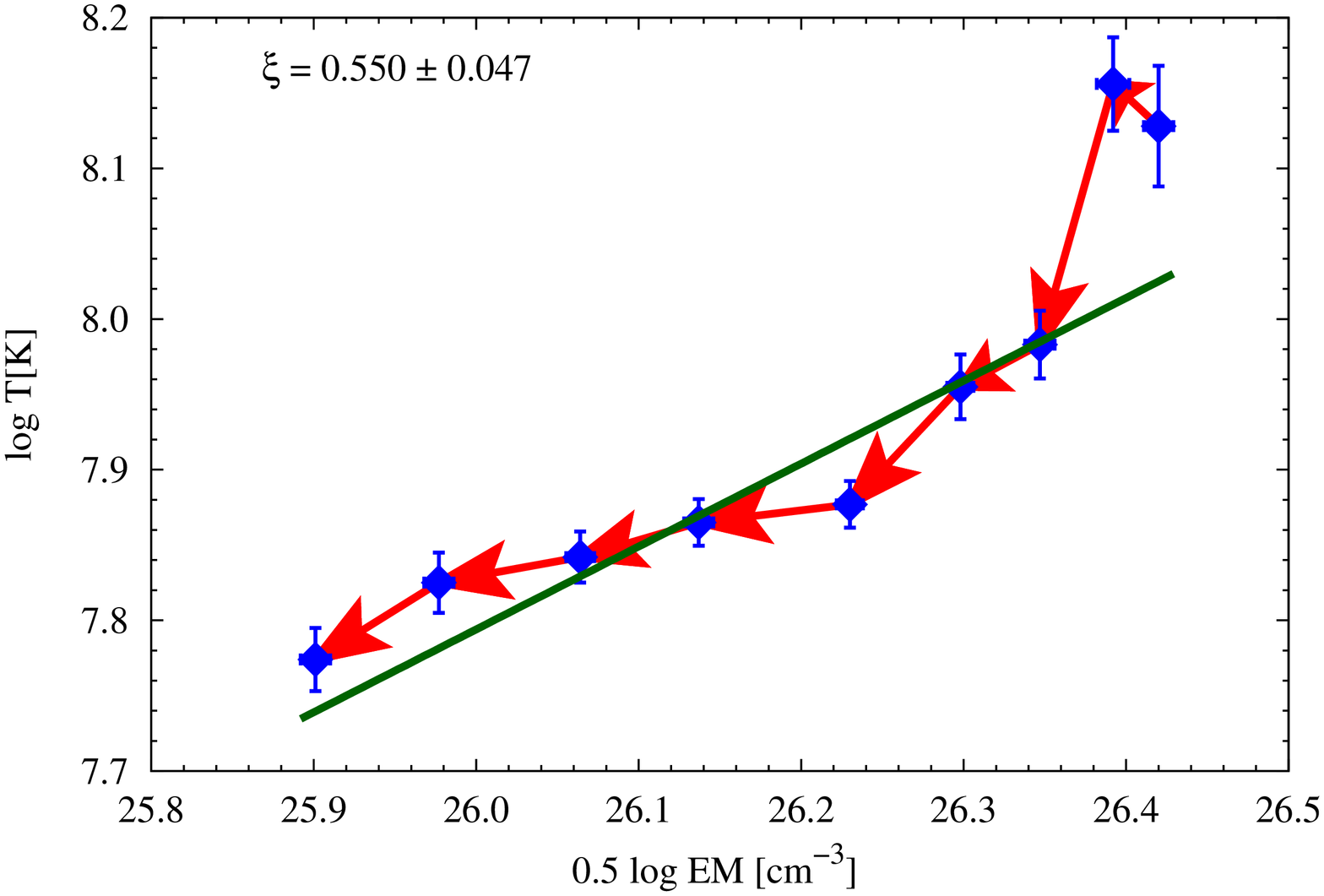}
\includegraphics[width=6cm,angle=-90]{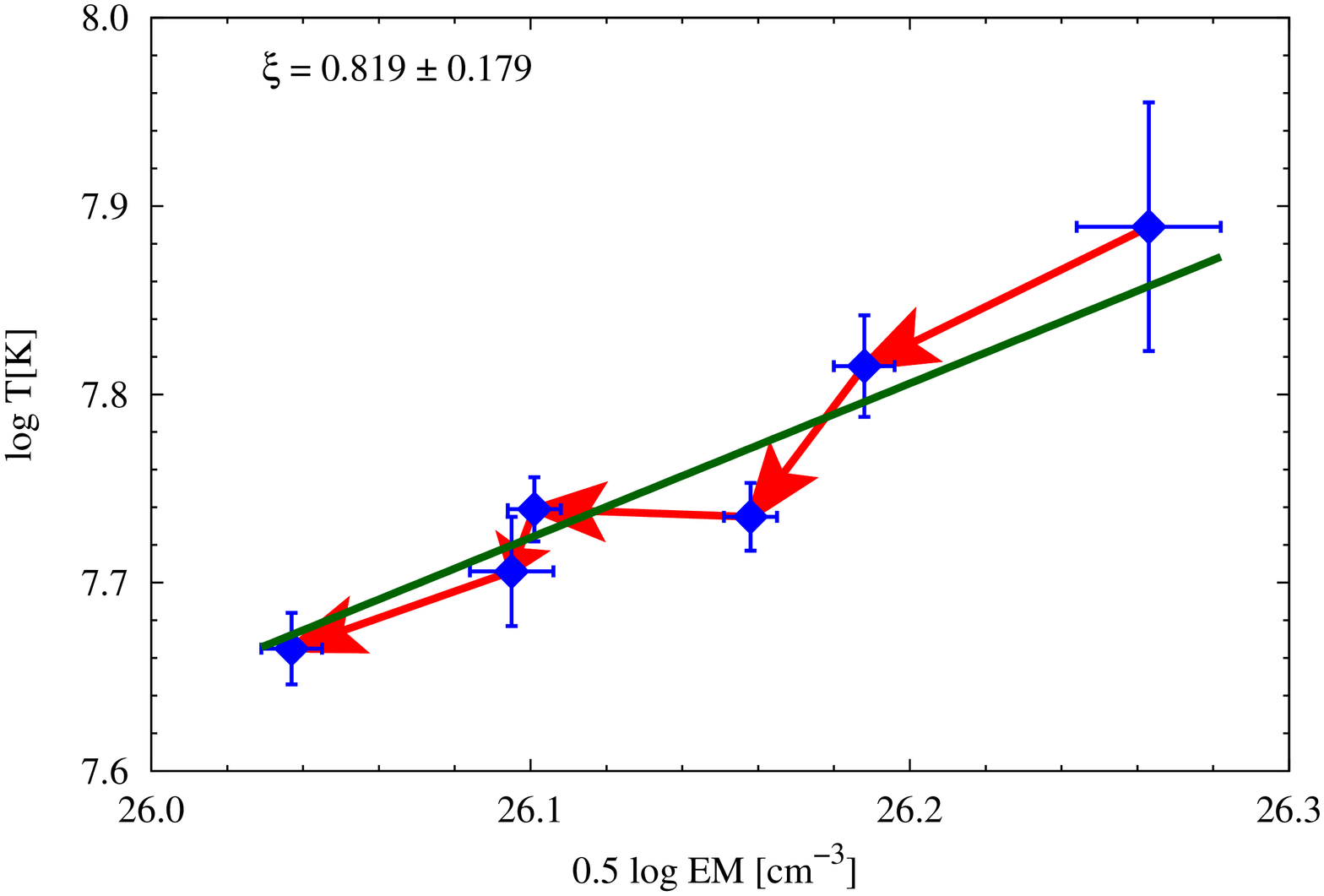}
\caption{Evolution of flares F1 (left) and F2 (right) in log $\sqrt{\rm EM}$ \textendash ~log T plane. The continuous line shows the best-fit during the decay phase of the flares with a slope $\zeta$ shown in the top left corner of each plot.}
\label{fig:zeta}
\end{figure*}

\begin{figure*}[t]
\includegraphics[width=11cm,angle=-90]{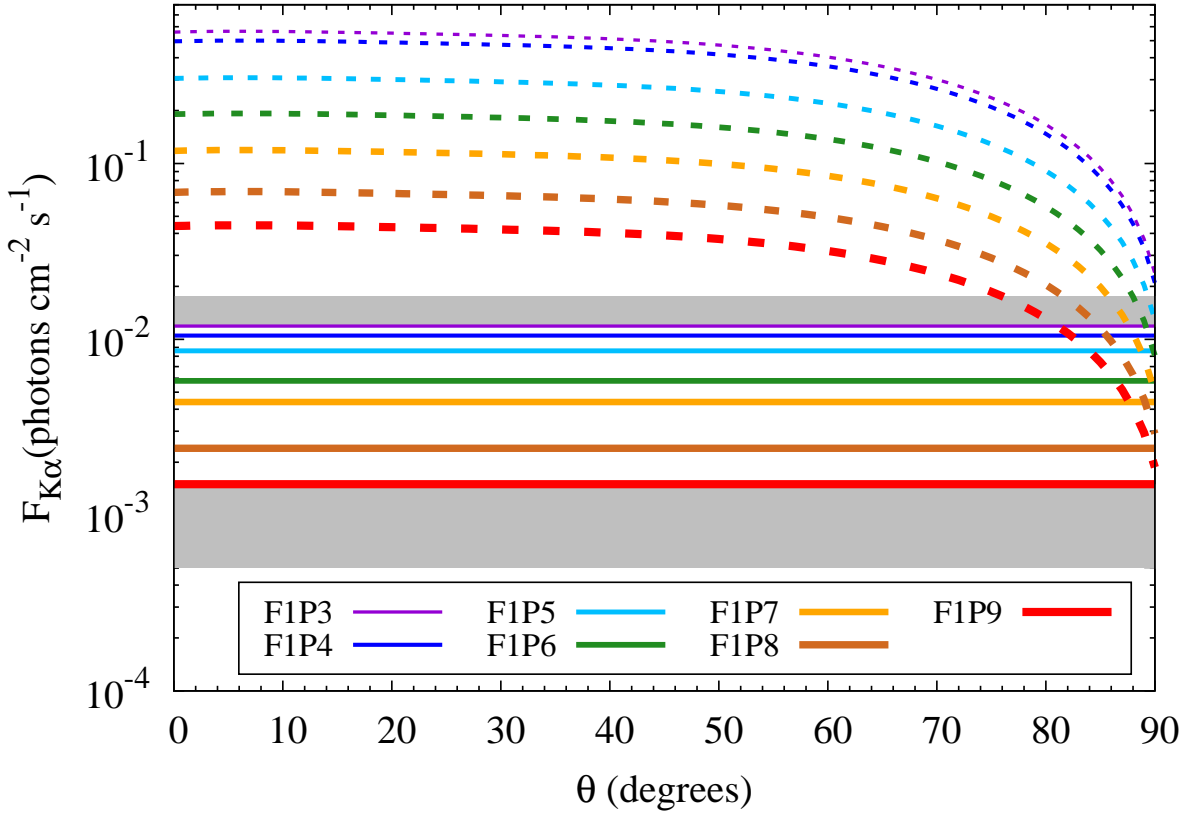}
\caption{Modeling of 6.4 keV line flux for the seven time intervals during the decay of flare F1. The solid horizontal lines show the observed Fe \Ka ~line fluxes of different time segments marked in the bottom of the plot. The dashed curves with similar thickness correspond to the modeled Fe \Ka ~line flux variation with the astrocentric angle for the same time intervals. Thicker the lines correspond to the later time spans. The dark shaded regions indicate the upper and lower 68\% confidence intervals of the first and last time segments, respectively. }
\label{fig:xrt_kapha-1_f1}
\end{figure*}

\subsection{Hydrodynamic modeling of flare decay}
\label{sec:loop-modeling}
Although stellar flares cannot be spatially resolved, it is possible to infer the physical size and structure of flares from the flare loop models. 
A time-dependent one-dimensional hydrodynamic model developed by \cite{Reale-97-1} to analyze the stellar flares assumes that the energy release occurs inside closed magnetic coronal structures, which confine the flaring plasma. 
The heat pulse is assumed to be released at the beginning of the flare at the apex of the flaring loop, and the thermodynamic cooling time scales the length of the flaring loop. Since the flaring plasma is also subjected to further prolonged heating, which extends into the decay phase of the flare, the actual flare decay time determined from the flare light curve would be longer than usual thermodynamic cooling time.  In the case of spatially resolved solar flares, \cite{Sylwester-93-2} showed that the slope ($\zeta$) in the density-temperature plane of the flare decay path provides a diagnostics of sustained heating. 
Using hydrodynamic simulations of semi-circular flaring loops with constant cross-sections, and including the effect of the heating in the decay, \cite{Reale-97-1} derived an empirical formula of loop length as
\begin{equation}
  L = \frac{\tau_{d}\sqrt{T_\mathrm{max}}}{1.85 \times 10^{-4} F(\zeta)}~~~~~~~~~~F(\zeta) \ge 1
 \label{eq:looplength}
 \end{equation}
where $L$ is the loop length in centimeters, $T_\mathrm{max}$ is the maximum temperature (K) of the loop apex, and $F(\zeta)$ is a \mbox{non-dimensional}
factor that provides a quantitative diagnostic of the ratio between the observed decay time ($\tau_{d}$) to the thermodynamic cooling time ($\tau_{th}$).
The actual functional form of the parameter $F(\zeta)$ depends on the bandpass and spectral response of the instruments. For \swift\ XRT observations, the $F(\zeta)$ was calibrated by \cite{Osten-10-5} as  
\begin{equation}
{\tau_{\rm d} \over \tau_{\rm th}} = F(\zeta) = {{1.81} \over {\zeta - 0.1}} + 0.67~~~~~~~~~~0.4 < \zeta \lesssim 1.9
\label{eq:fzeta}
\end{equation}
This limits the applicability of this model, because it assumes a flare process as a sequence of quasi-static states for the loop, in which the heating time-scale ($\tau_{\rm H}$) is very long in order to mask the loop's intrinsic decay  ($\tau_{\rm H} \gg \tau_{\rm th}$) in the lower end, and a freely decaying loop with no heating ($\tau_{\rm H} = 0$) on the upper end.
For stellar observations, no density determination is normally available; therefore, we used the quantity $\sqrt{\rm EM}$ as a proxy of the density assuming the geometry of the flaring loop during the decay. Fig.~\ref{fig:zeta} shows the path \mbox{log $\sqrt{\rm EM}$ versus log $T$} for flares F1 and F2.
A linear fit to the data provided the value of $\zeta$ as $0.550 \pm 0.047$ for flare F1 and $0.819 \pm 0.179$ for flare F2. This indicates the presence of sustained heating during the decay phase of both flares. 

The relationship between  $T_\mathrm{max}$ and observed peak temperature ($T_\mathrm{obs}$) was also calibrated for \swift\ XRT by \cite{Osten-10-5} as $T_{\rm max} = 0.0261 \times T_{\rm obs}^{1.244}$, where both temperatures are in K. For flares F1 and F2, $T_{\rm max}$ was  calculated to be 365$\pm$33 and 170$\pm$32 MK, respectively. 
The flaring loop length was derived as 1.2$\pm$0.1\E{10} cm for flare F1 and 2.2$\pm$0.6\E{10} cm for flare F2. Assuming a semi-circular geometry, the flaring loop height (L/$\pi$) was estimated to be 0.1 and 0.2 times the stellar radius (\rstar) of the primary component of CC Eri of the flares F1 and F2.
The loop parameters for both the flares are given in Table~\ref{tab:par_loop}.

\subsection{Energetics}
\label{sec:energy}
The derived loop lengths are much smaller than the pressure scale height\footnote{The pressure scale height is defined as $h_\mathrm{p} = 2kT_\mathrm{max}/(\mu g)$, where $\mu$ is the average atomic weight and $g$ is the surface gravity of the star. Considering both the stellar components, the derived values of $h_\mathrm{p}$ are $\ge$9.8\E{11} cm for flare F1 and $\ge$4.6\E{11} cm for flare F2.} of the flaring plasma of CC Eri. Therefore, we can assume that the flaring loop is not far from a steady-state condition. We applied the RTV scaling laws \citep{Rosner-78-4} to determine the maximum pressure (p) in the loop at the flare peak and found it to be $\sim$1.5\E{6} and $\sim$8\E{4} dyne cm$^{-2}$ for the flares F1 and F2, respectively. 
The plasma density ($n_e$), the flaring volume ($V$),  and the minimum magnetic field ($B$) to confine the flaring plasma were derived as

\begin{eqnarray}
n_e = \frac{p}{2 k T_\mathrm{max}}~{\rm cm^{-3}};~
V = \frac{\rm EM}{n_e^{2}}~{\rm cm^3}; B = \sqrt{8\pi p}~{\rm G}
\end{eqnarray}
 \label{eq:params}

 The estimated values of  $n_e$, $V$, and $B$ during the flares F1 and F2 were 1.5\E{13}  and  1.7\E{12} cm$^{-3}$, 3.1\E{28} and 1.2\E{30} cm$^3$, and $\sim$6.1 and $\sim$1.4 kG, respectively. 
 Using the RTV scaling laws, we have also estimated the heating rate per unit volume ($E_H = \frac{d H}{d V d t} \simeq 10^5 ~ p^{7/6} ~ L^{-5/6}$) at the peak of the flare as $\simeq$6.6\E{3} and $\simeq$1.3\E{2} $\rm erg ~ cm^{-3} ~ s^{-1}$ for the flares F1 and F2, respectively.
The total heating rates ($ \frac{d H}{dt} \simeq \frac{d H}{d V d t} \times V$) at the peak of the flares were derived to be $\sim$2.1\E{32} and $\sim$1.5\E{32} $\mathrm{erg~s^{-1}}$ for the flares F1 and F2, which were, respectively, $\sim$1.2 and  $\sim$2.5 times higher than the flare maximum luminosity.
If we assume that the heating rate is constant throughout the rise and decay phases of the flare, the total energy radiated [$E_{\rm X,tot} > \frac{d H}{dt} \times (\tau_r + \tau_d)$]  during the flares was derived to be $>$1.4\E{35} erg for flare F1 and $>$1.7\E{35} erg for flare F2.  These values were $\sim$40 s and $\sim$47 s of bolometric energy output of the primary and  $\sim$221 s and $\sim$264 s of bolometric energy output of the secondary for the flares F1 and F2, respectively.

\subsection{Modeling of Fluorescent Fe \Ka\ Emission}
\label{sec:fe-modeling}
In the stellar context, the detected iron \Ka\ line is generally attributed to a fluorescent process, where the fluorescing material is a  neutral or low ionization state of photospheric iron (\ion{Fe}{1}--\ion{Fe}{12}), which shines on the X-ray continuum emission arising from a loop-top source. Thus the detection of this line constraints the height of the flaring loop. The process involves photoionization of an inner K-shell electron and the de-excitation of an electron from a higher level at this energy. 
Thus the total photon flux above the Fe \Ka\ ionization threshold of 7.11 keV is one of the main contributors to the observed flux in the Fe \Ka\ line. For the solar flares, \cite{Bai-79-3} derived  a formula for the flux of Fe \Ka\ photons received on the Earth, which was later extended to stellar context by \cite{Drake-08-35} and is given by 
\begin{equation}
  F_{K\alpha} = f(\theta) \Gamma(T,h) F_{7.11} \;\;\;{\rm photons \; cm^{-2} \; s^{-1}}
\end{equation}
where $F_{7.11}$ is the total flux above 7.11 keV, $f(\theta)$ is a function
that describes the angular dependence of the emitted flux on the astrocentric angle (defined as an angle subtended by the flare and the observer), and $\Gamma$ is the fluorescent efficiency. We used the coefficients derived by \cite{Drake-08-35} to determine the functional dependence of $f(\theta)$ and  $\Gamma$ for different loop heights \citep[see Tables 2 and 3 of][]{Drake-08-35}. We could only get enough statistics in both XRT and BAT spectra (required to estimate $F_{7.11}$) for the flare F1; therefore, we have done our analysis only for flare F1. 
The value of $\Gamma$ was taken as 0.96 for a loop height of 0.1 \rstar\ and a temperature of 100 MK, which are closest to the derived loop length and maximum temperature of the flare F1.
Because photospheric iron abundances of CC Eri are not known, a default value of abundances of  3.16\E{-5}\  was used in the calculations \cite[see][]{Drake-08-35}.
Since the flare F1 was only detected up to 50 keV (see \S~\ref{sec:xray-lc}), in our analysis, we considered the upper limit of energy to be 50 keV. The model flux of the photon density spectrum above 7.11 keV was calculated by using best-fit \apec\ model parameters from the joint spectral fitting of XRT+BAT spectra in different time intervals for flare F1.
Fig.~\ref{fig:xrt_kapha-1_f1} shows the modeled Fe \Ka\ flux as a function of the astrocentric angle of a flare height of 0.1 \rstar\ for different time segments. The corresponding observed flux is also shown by continuous lines of the same thickness. The modeled and observed Fe \Ka\ line flux was found to overlap at an astrocentric angle of $\sim$90\deg.

\section{Discussion and Conclusions}
\label{sec:discussions}
\subsection{Temporal and spectral properties}
\label{sec:disc-params}

In this paper, we have presented a detailed study of two X-ray superflares observed on an active binary system CC Eri with the \swift\ satellite. These flares are remarkable in the large enhancement of peak luminosity in soft and hard X-ray energy bands. A total of eight flares have been detected in X-ray bands on CC Eri thus far. Out of the eight flares, two flares are the strongest flares in terms of energy released.
The soft X-ray luminosity increased up to $\sim$400 and $>$3 times more than to that of the minimum observed values for the flares F1 and F2, respectively. The former is much larger than any of the previously reported flares on CC Eri observed with \textit{MAXI} GCS \citep[$\sim$5--6 times more than quiescent;][]{Suwa-11-1}, \chandra\ \citep[$\sim$11 times more than quiescent;][]{Nordon-07}, \xmm\ \citep[$\sim$2 times more than quiescent;][]{Crespo-Chacon-07, Pandey-08-4}, \rosat\ \citep[$\sim$2 times more than quiescent;][]{Pan-95-9},  and other flares observed with \textit{EXOSAT} \citep{Pallavicini-88-1}, \textit{Einstein} IPC \citep{Caillault-88-3}, and \textit{HEAO1} \citep{Tsikoudi-82}. However, similar magnitude flares have been reported in other stars such as DG CVn \citep{Fender-15-5,Osten-16-1}, EV Lac \citep{Favata-00-6, Osten-10-5}, II Peg \citep{Osten-07-3}, UX Ari \citep{Franciosini-01-1}, AB Dor \citep{Maggio-00}, Algol \citep{Favata-99-1}, and EQ1839.6+8002 \citep{Pan-97-7}.
Considering that the flare happened in the primary (K7.5 V), the peak X-ray luminosity of flare F1 and flare F2 were, respectively, found to be 48\% and 16\% of bolometric luminosity (\lbol); if the flare happened in the secondary (M3.5 V) star, the peak X-ray luminosity of flares F1 and F2 were 267\% and 91\% of \lbol, whereas the peak luminosity for flares F1 and F2 were found to be 41\% and 14\% of combined \lbol, respectively.

Both  flares  appear to be the shortest duration flares observed on CC Eri thus far. However, a weak flare with a similar duration was observed by the \xmm\ satellite \citep{Crespo-Chacon-07}. The durations of all other previously observed flares on CC Eri were in the range of 9--13 ks. The durations of the superflares on CC Eri are also found to be smaller than other observed superflares, e.g. $\sim$3 ks for EV Lac \citep{Osten-10-5}, $>$10 ks for II Peg \citep{Osten-07-3}, $\sim$14 ks for AB Dor \citep{Maggio-00}, and $\sim$45 ks for Algol \citep{Favata-99-1}.
The e-folding decay times of both X-ray flares are shorter in the hard spectral band than those in the softer band.

During the flares F1 and F2, the observed temperature reached a maximum value of  $\sim$174 MK for flare F1 and $\sim$128 MK, respectively.
These values of temperatures are quite high from previously observed maximum flare temperatures on CC Eri \citep{Crespo-Chacon-07, Nordon-07, Pandey-08-4}, but  are of the similar order to those of other superflares detected on II Peg \citep[$\approx$300 MK;][]{Osten-07-3}, DG CVn \citep[$\approx$290 MK;][]{Osten-16-1}, EV Lac \citep[$\approx$150 MK and $\approx$142 MK;][]{Favata-00-6, Osten-10-5}, and AB Dor \citep[$\approx$114 MK;][]{Maggio-00}.  
The abundances during the flares F1 and F2 are found to enhance $\sim$9 and $\sim$2 times more than those of the minimum values observed. During other superflares, abundances were found to increase between two to three times more than that of the quiescent level \citep{Favata-99-1, Favata-00-6, Maggio-00}. However, in the case of the superflare observed with \swift\ in EV Lac, the abundances were found to remain constant throughout the flare. From Fig.~\ref{fig:trs_batxrt}, it is evident that the abundance peaks after the temperature and luminosity peaks, which is consistent with a current idea in the literature \citep[see][]{Reale-07-3}. This could be due to the heating and evaporation of the chromospheric gas, which increases the metal abundances in the flaring loop.
For both flares, the temperature was peaked before the EM did. A similar delay was also observed in many other solar and stellar flares \citep[e.g.][]{Sylwester-93-2, Favata-99-1, Favata-00-6, Pandey-08-4}. The temperature increases due to beam driven plasma heating and later subsequent evaporation of the plasma into upper parts of the coronal loop that increases its density, and therefore EM ($\sim$n$_e^{2}$). Later coronal plasma cools down by thermal conduction and then via radiative losses
\citep[e.g.][]{Cargill-04-3}.

\subsection{Coronal loop properties}
\label{sec:disc-loop}

The derived loop lengths for the flares F1 and F2 are larger than  previously observed flares by \cite{Crespo-Chacon-07} on CC Eri.
The loop lengths are also in between the loop lengths derived for other G--K dwarfs, dMe stars, and RS CVn type binaries \citep[e.g.][]{Favata-03-1, Pandey-08-4, Pandey-12-19}.
Instead of $\sim$3 times larger peak luminosity in flare F1 than that of the flare F2, the derived  loop length for flare F2 is two times larger than that of flare F1, which might be interpreted as a result of an $\sim$1.4 times more sustained heating rate in the decay phase of flare F1 than that of flare F2. 
The heating rate during flare F1 is also found to be $\sim$49\% of the bolometric luminosity of the CC Eri system, whereas during flare F2 the heating rate is only $\sim$35\% of the combined bolometric luminosity. 
The derived heating rate is also found to be more than the maximum X-ray luminosity for both flares.  This result is compatible with X-ray radiation being one of the major energy loss terms during the flares.

Present analysis allows us to make some relevant estimation of the magnetic field strength that would be required to accumulate the emitted energy and to keep the plasma confined in a stable magnetic loop configuration. Under the assumptions that the energy release is indeed of
magnetic origin, the total non-potential magnetic field $B_{\rm 0}$ involved in a flare energy release within an active region of the star can be obtained from the relation
\begin{center}
  \begin{equation}
    \label{eq:b}
    E_{\rm X,tot} = {{(B_{\rm 0}^2-B^2)} \over 8 \pi}\times V
  \end{equation}
\end{center}
Assuming that the loop geometry does not change during the flare, $B_{\rm 0}$ is estimated to be $>$12 and $>$2 kG for the flares F1 and F2, respectively. \cite{Bopp-73-4} also estimated a large magnetic field of 7 kG on CC Eri at photospheric level.
We have used the loop volume in the derivation of $B_{\rm 0}$, but  this may not imply that the magnetic field fills up the whole volume. Rather, our estimation of $B_{\rm 0}$ is based on the assumption that the energy is stored in the magnetic field configuration (e.g. a large group of  spots) of the field strength of  several kG with a volume comparable to one of the flaring loops.
 
\subsection{Flare Location and Fe \Ka\ emission feature}
\label{sec:disc-ka}
Given that CC Eri is an active binary system, we can consider three possible scenarios of the flare origin: (i) energy release occurs due to magnetic reconnection between magnetic fields bridging two stars \citep[see][]{Uchida-83-1, Graffagnino-95}, (ii) flares occurred on K-type primary star, and (iii) flares occurred on M-type secondary star.  The binary separation of the CC Eri system of 1.4 \E{11} cm \citep{Amado-00-1, Crespo-Chacon-07} is more than an order of the height of flaring loops  for  both flares. 
Therefore, it is more likely that the flares are attached to a corona of any one of the components of CC Eri. It is also very difficult to identify the component of the binaries on which the flares occured. 

One of the most interesting findings is the detection of Fe \Ka\ emission line in the X-ray spectra of CC Eri during the flares, whose flux depends on the photospheric iron abundance, the height of the emitting source, and the astrocentric angle between the emitting source and observer\citep{Bai-79-5,Drake-08-35}.
Recently, the Fe \Ka\ emission line was also detected in several other cool active stars during large flares, such as HR 9024 \citep{Testa-08-6} and II Peg \citep{Ercolano-08-1}, and has been well described by the fluorescence hypothesis. In most of the cases, the flaring loop length derived in this method was found to be consistent with the loop length derived from the hydrodynamic method.
Using the loop length derived from the hydrodynamic model, the astrocentric angle between the flare and observer has been estimated as $\sim$90\deg. This shows that the region being illuminated by the flare, and thus fluorescing the photospheric iron, is located near the stellar limb.

\acknowledgments
This work made use of data supplied by the UK Swift Science
Data Centre at the University of Leicester. This research has made use of the
  XRT Data Analysis Software (XRTDAS) developed under the responsibility
  of the ASI Science Data Center (ASDC), Italy. We thank the anonymous referee for useful comments.

{\it Facility:} \facility{\swift ~(BAT, XRT)}.

\bibliography{SK_collections}{}
\end{document}